\documentclass[a4paper]{article}

\usepackage[utf8]{inputenc} % allow utf-8 input
\usepackage{hyperref}       % hyperlinks
\usepackage{url}            % simple URL typesetting
\usepackage{amsfonts}       % blackboard math symbols
\usepackage{amsmath}
\usepackage{graphicx}
\usepackage{caption}
\usepackage{float}
\usepackage[left=3cm,right=3cm,top=2cm,bottom=2cm]{geometry}
\usepackage[english]{babel}
\usepackage{amsmath}
\usepackage{graphicx}
\usepackage[colorinlistoftodos]{todonotes}
\usepackage{float}
\usepackage{dsfont}
\usepackage{bm} % used for \boldsymbol
\usepackage{booktabs} % Horizontal rules in tables
\usepackage{multicol}
\usepackage{multirow}
\usepackage{gensymb}
\usepackage{tabularx}
\usepackage{cite}
\usepackage[normalem]{ulem}
\DeclareCaptionFormat{sanslabel}{#3}%

\begin{document}

\Large
\begin{center}
\vspace{1cm}
\textbf{Adaptive-glasses wavefront sensorless full-field OCT for high-resolution retinal imaging over a wide field-of-view}

\vspace{0.5cm}

% Author names and affiliations
\normalsize
Jules Scholler$^{1,2}$, Kassandra Groux$^{1,2}$, Kate Grieve$^{2}$, Claude Boccara$^{1,2}$ and Pedro Mecê$^{1,2}$

\end{center}
\scriptsize
$^1$Institut Langevin, ESPCI Paris, CNRS, PSL University, 1 rue Jussieu, Paris, France\\
$^2$Paris Eye Imaging Group, Quinze Vingts National Ophthalmology Hospital, Paris, France\\
$^*$pedro.mece@espci.fr

\normalsize

%%%%%%%%%%%%%%%%%%% abstract %%%%%%%%%%%%%%%%
%% [use \begin{abstract*}...\end{abstract*} if exempt from copyright]

\begin{abstract}
The highest three-dimensional (3D) resolution possible in \textit{in-vivo} retinal imaging is achieved by combining optical coherence tomography (OCT) and adaptive optics (AO). However, this combination brings important limitations, such as small field-of-view and complex, cumbersome systems, preventing so far the translation of this technology from the research lab to clinics. In this Letter, we introduce an approach that avoids these limitations by using a multi-actuator adaptive lens just in front of the eye, in a technique we call the adaptive-glasses wavefront sensorless approach. We implemented this approach on our compact full-field OCT (FFOCT) retinal imager without increasing its footprint or optical complexity. The correction of ocular aberrations through the adaptive-glasses approach increased the FFOCT signal-to-noise ratio and enabled us to image different retinal layers with a 3D cellular resolution in a $5\degree \times 5\degree$ field-of-view, without apparent anisoplanatism.  
\end{abstract}

%%%%%%%%%%%%%%%%%%%%%%%%%%  INTRO  %%%%%%%%%%%%%%%%%%%%%%%%%%
\section{Introduction}
Optical coherence tomography (OCT) revolutionized ophthalmology in the 1990s owing to its high axial resolution, which enabled clinicians to distinguish the retinal layers \textit{in-vivo} \cite{huang_optical_1991}. Although the axial resolution of OCT is sufficient to resolve retinal features at a micrometer scale, the lateral resolution is limited by ocular aberrations \cite{Jarosz_aberrometry_50_17}. Owing to its capacity to correct for ocular aberrations in real-time, adaptive optics (AO) has become the primary technique to achieve high lateral resolution in the retina \cite{liang1997supernormal}. When coupled to OCT, AO has enabled micrometer resolution in all spatial dimensions for \textit{in-vivo} retinal imaging \cite{jonnal2016review}, contributing to the understanding of retinal function \cite{azimipour2019functional} and diseases \cite{lassoued2020dysfunction}. Nevertheless, the high lateral resolution achieved with AO comes with a cost of a small field-of-view (FOV), which is limited by the isoplanatic patch of the eye (around $2\degree \times 2\degree$) \cite{Bedggood_isoplanetic_patch_2008} but also by the trade-off between the spatial sampling of the scan and the acquisition speed, in order to avoid image distortion due to eye motion. Most importantly AO-OCT systems are complex and cumbersome, requiring long imaging sessions to acquire a large FOV \cite{jonnal2016review}. These limitations have prevented the translation of AO-OCT from the research lab to clinics. Much recent work has aimed at addressing these limitations, by reducing the AO-OCT system complexity and footprint through the use of conjugated lens-based wavefront sensorless AO \cite{jian2016lens,Verstraete_DONE_17}, or by increasing the field-of-view to $4\degree \times 4\degree$ using multi-conjugate AO \cite{brunner2020single}. However, the former still presents a small FOV because of the limited isoplanatic patch of the eye and spatial sampling of the scan, and the latter adds complexity as two deformable mirrors are necessary.

In this Letter, we propose a lens-based sensorless AO approach using a multi-actuator adaptive lens (MAL - Dynamic Optics srl, Italy) positioned in front of the examined eye, \textit{i.e.} without strict pupil conjugation, in a technique we call the adaptive-glasses approach. Since the MAL has a larger diameter than the eye's pupil, positioning it close to the eye allows an anisoplanatic correction, therefore increasing the FOV over which AO is effective, at the cost of a slight decrease in the maximum spatial frequency of the applied wavefront correction \cite{Mertz_conjuguate_AO_2015}. The proposed optimization scheme does not require any calibration step and is therefore straightforward to implement in existing systems without increasing system footprint or optical complexity. We implemented the adaptive-glasses approach in our compact time-domain full-field OCT (FFOCT) system. FFOCT was found to behave differently to conventional imaging systems and Fourier-domain OCT with regards to optical aberrations. The lateral resolution of FFOCT is less affected by symmetric aberrations \cite{Tricoli_19,barolle2019approche}, which dominate in the eye, owing to the use of a spatially incoherent source. Nevertheless, although symmetrical aberrations may not be adversely affecting the resolution in FFOCT, the presence of aberrations still provokes a loss of signal-to-noise ratio (SNR), as they dampen spatial frequencies carrying structural information. Recently, non emmetropic subjects were imaged with FFOCT while wearing their prescription eyeglasses, which increased SNR for photoreceptor imaging  \cite{Mece_axial_stab_20}. Prescription eyeglasses are limited to correcting only low order aberrations with low precision ($\pm 0.125D$ \cite{Thibos_eye_aberration_02}) however, impacting the robustness of FFOCT imaging, especially at large pupil size, and preventing imaging of retinal layers other than photoreceptors \cite{Mece_axial_stab_20}. Here, we show that the use of the adaptive glasses approach to correct for ocular aberrations can considerably increase the SNR and robustness of FFOCT, enabling imaging of different retinal layers with 3D cellular resolution over a $5\degree \times 5\degree$ FOV acquired in a single shot, while retaining a compact system design. To the best of our knowledge the presented method is the first to ally 3D high-resolution, wide FOV and small system footprint, which are essential characteristics for clinical deployment.

%%%%%%%%%%%%%% Methods %%%%%%%%%%%%%%%%%%%%%%%
\section{Methods}
\subsection{Experimental set-up}
%% FF-OCT setup

The FFOCT system was described in detail elsewhere \cite{Mece_axial_stab_20} and has a footprint of $50 cm~ \times~ 30 cm$ (See \textcolor{blue}{Visualization 1} for a mechanical drawing of the system). Importantly for this study, the FFOCT setup comprises a spectral-domain (SD) OCT channel used to track eye axial motion and drive a fast translation stage on which the FFOCT reference arm is mounted, enabling the FFOCT to acquire \textit{en-face} images at a given depth in the retina. Since the brightness of FFOCT retinal images varies with phase modulation, the FFOCT signal itself cannot be used as a merit function for wavefront correction. We therefore use the brightness of the SD-OCT B-scan as a surrogate for the FFOCT SNR optimization. In addition, SD-OCT B-scans are also faster to acquire and process than FFOCT images, benefiting rapid wavefront optimization. The FFOCT and SD-OCT channels have central wavelengths of 850nm (30nm bandwidth) and 930nm (60nm bandwidth) respectively. Since these two wavelengths are close, correcting aberrations using the brightness of SD-OCT as a merit function is also suitable for FFOCT. % Having different but close wavelengths is a common practice in AO for retinal imaging, one used as a laser beacon for the wavefront sensor and the other one for imaging \cite{gofas2018high}. 
%\begin{figure}[H]
%    \centering
%    \includegraphics[width= 0.7\linewidth]{Setup.png}
%    \caption{Schematic drawing of the custom-built FF-OCT system coupled with an SD-OCT. \textcolor{blue}{Visualization 1} shows an animation of the opto-mechanical drawing of the present system with a footprint of $50 cm \times 25 cm$.}
%    \label{fig:UsafTarget}
%\end{figure}
For both FFOCT and SD-OCT channels, the beam diameter at the pupil is 7.5mm. The MAL is composed of 18 actuators. It has a transmission of 94\% in the near infrared and a response time of less than 2ms \cite{Verstraete_DONE_17}. The MAL is positioned 2-3 cm in front of the subject's cornea, \textit{i.e.} without strict pupil conjugation, and has a 10mm diameter, meaning that it is large enough to avoid vignetting and resolution loss which would occur with a smaller numerical aperture, and favoring anisoplanatic correction through wavefront sensorless optimization \cite{Mertz_conjuguate_AO_2015}. The coherence gate geometry was shaped to fit the retinal curvature and dispersion was compensated using a 20mm N-BAK1 optical window in the sample arm \cite{mece_curved_field_2020}. The size of an individual pixel of the FFOCT camera corresponds to $1\mu m$ in the retinal plane. The SD-OCT has an A-scan rate of 36kHz. We chose to scan over a $2\degree$ FOV with 256 A-scans, \textit{i.e.} at 140 Hz, providing a good trade-off between acquisition speed and SNR \cite{Mece_axial_stab_20}. B-Scans were averaged in the lateral dimension, and used for three purposes: 1) tracking the eye axial motion for correction in real-time, 2) guiding positioning of the FFOCT coherence gate at the layer of interest, and 3) as a merit function for the wavefront optimization. The merit function can be applied for any retinal layer of interest to automatically adjust the MAL shape to favor imaging in that particular layer.

\subsection{Wavefront sensorless approach}

\begin{figure}[H]
    \centering
    \includegraphics[width=0.6\linewidth]{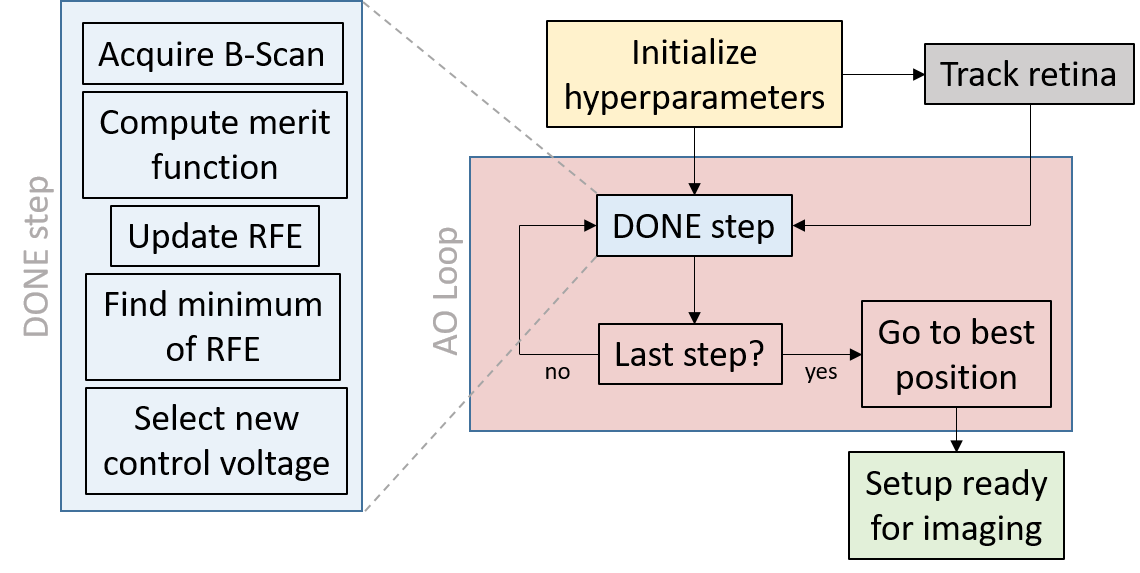}
    \caption{The retina is tracked throughout the optimization to compute the merit function on the selected layer.The number of iterations was set to 50, for a total of 1s optimization duration.}
    \label{fig:AO_diagram}
\end{figure}

To perform the optimisation, we used the DONE algorithm \cite{Bliek_DONE_2018} which had three major advantages for the problem we wished to solve. First, it does not require evaluation of the merit function gradient, hence mitigates the MAL hysteresis by limiting the number of times that voltages are updated. Second, the DONE algorithm is very stable due to the use of regularization and is therefore well suited for noisy experimental data. Finally, as we use the actuator voltages as input degree of freedom, no calibration step is necessary since no modal decomposition is used. The DONE algorithm models the unknown merit function using a Random Fourier Expansion (RFE) $g(x) = \sum_{k=1}^D c_k cos(\omega_k^T x +b_k)$ fit to the experimental data using a least square approach. It iteratively finds a minimum of the merit function on a compact set $X \subseteq [V_{min}, V_{max}]^d$ representing each actuator voltage (where $d=18$ is the number of actuators) by updating the RFE at each new measurement, and using this RFE as a surrogate of the merit function for optimization. Hyperparameters were selected by trial and error by imaging a model eye, and then refined for \textit{in-vivo} imaging. The number of basis functions D was set to 100. Increasing D leads to a better RFE fit at the cost of more computation time (complexity is $O(D^2)$). In order to prevent underfitting and overfitting, a regularization parameter $\lambda$ is used in the least square fit for finding the RFE coefficients $c_k$. It also helps with dealing with few measurements. We set $\lambda= 0.01$. Finally, the probability density function of frequencies $\omega$ for the RFE model are drawn from a Gaussian distribution (variances $\sigma_\zeta^2=\sigma_\xi^2=1$) and influences the exploration of the RFE surrogate and the original function, respectively. The optimisation steps are described in Fig.~\ref{fig:AO_diagram}.

%%%

 \subsection{Image acquisition and processing}
%% Imaging acquisition and processing
Retinal imaging was performed on three healthy subjects. Research procedures followed the tenets of the Declaration of Helsinki. Informed consent was obtained from subjects after the nature and possible outcomes of the study were explained. The study was authorized by the appropriate ethics review boards (CPP and ANSM (IDRCB number: 2019-A00942-55)). Each subject was seated in front of the system and stabilized with a chin and forehead rest and asked to fixate a target. To maximize pupil diameter, image acquisition was performed in a dark room. Images were acquired just after wavefront optimization in parallel with real-time correction of axial eye motion. As the DONE algorithm has a high robustness to noise, subjects could blink during the wavefront optimization routine. Image sequences were composed of 150 frames acquired at 300 Hz. The FFOCT camera worked close to saturation in order to use the whole full well capacity, so decreasing the relative importance of shot noise \cite{scholler_probing_2019}. During image acquisition, the total power entering the eye from the FF-OCT illumination source and the SD-OCT scanning source were respectively 1.3~mW (for the 0.5~s of image acquisition) and 0.25~mW (continuous scanning), which are below the ocular safety limits established by the ISO standards for group 1 devices. Since the phase of FFOCT was randomly modulated by the residual tracking error of axial motion, and to eliminate the incoherent terms, we adopted a 2-phase demodulation \cite{Mece_axial_stab_20}. The 2-phase demodulation consists of subtracting one image $I_{N}$ from the next $I_{N+1}$ and taking the absolute value. Next, images with a very low or absent signal, mainly due to an insufficient phase shift between consecutive images due to residual eye motion, were automatically detected, and then excluded from the image sequence before registration and averaging. In addition, we used the ImageJ plugin MosaicJ \cite{thevenaz2007user} to stitch together five images into a $12\degree \times 12\degree$ FOV. A photoreceptor density map was computed using a fully automated algorithm based on modal spacing as described in \cite{cooper2019fully}. 
%%%%%%%%%%%%%%%%%%%%%%%%%%  Results and discussion %%%%%%%%%%%%%%%%%%%%%%%%%%
\section{Results and discussion}
\begin{figure}[H]
    \centering
    \includegraphics[width= 0.75\linewidth]{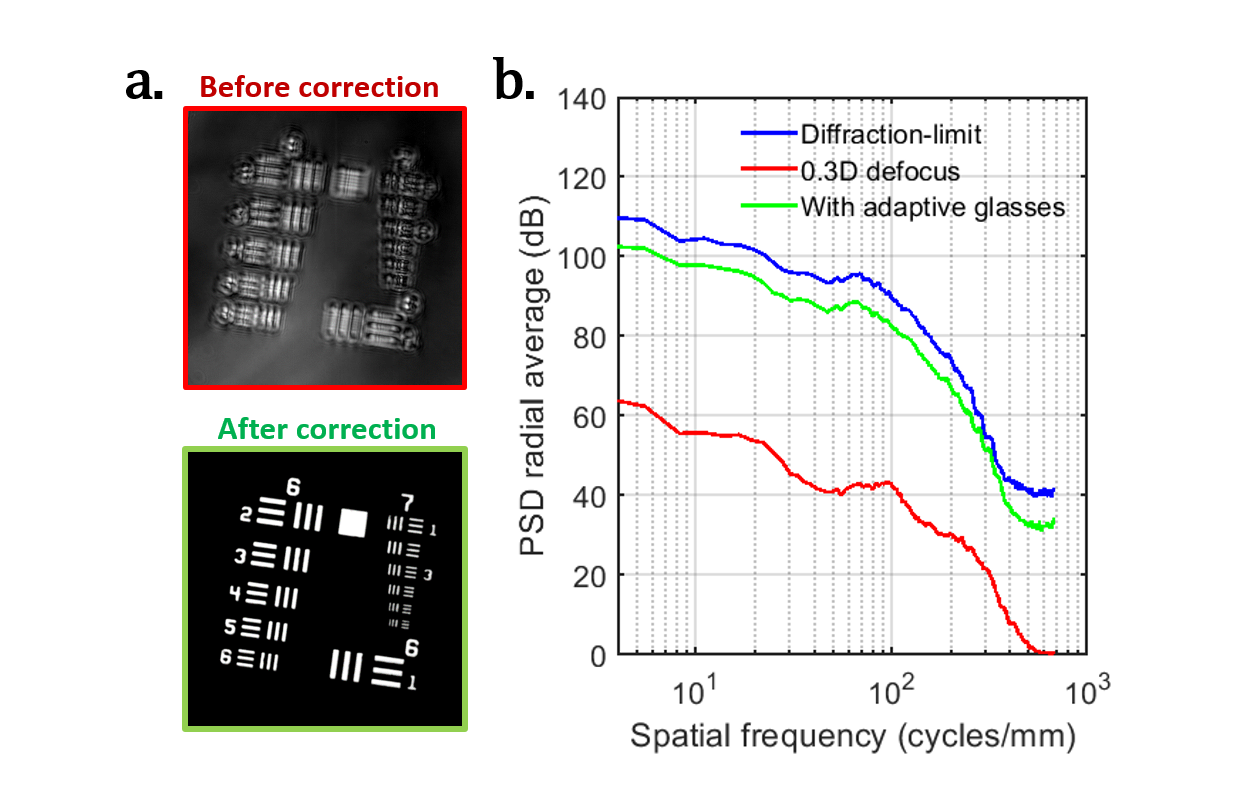}
    \caption{ \textbf{a.} USAF target image before and after aberration correction through the adaptive glasses approach using FFOCT. Although the SD-OCT brightness was used as a merit function, the \textit{en-face} images could also be used, though requiring longer processing time. \textbf{b.} PSD of USAF target images presented in \textbf{(a.)} compared to the expected diffraction limit case.}
    \label{fig:UsafTarget}
\end{figure}

We first tested the adaptive-glasses approach on a 1951 USAF target (Fig.~\ref{fig:UsafTarget}) after adding a microscope objective in the FFOCT sample arm and positioned the MAL at the back plane of the microscope objective. We added a 0.3D defocus in the sample arm and acquired FF-OCT images before and after correcting aberrations (Fig.~\ref{fig:UsafTarget}(a)). Aberrations induce phase artifacts yielding ringing effects and inversion of contrast. All of these were corrected after using the adaptive-glasses approach. Figure~\ref{fig:UsafTarget}(b) shows the power spectral density (PSD) of these images compared to the diffraction-limited PSD. Note that all three PSD plots present an almost identical distribution of spatial frequencies, showing the robustness of FF-OCT to symmetric aberrations (here defocus) in terms of resolution, whilst SNR drops significantly. Using the adaptive-glasses approach, we were able to recover the lost SNR, almost reaching the value we would expect for diffraction-limited imaging. 

%\textcolor{blue}{Visualizations 1,2} show the temporal evolution of the fringes contrast (responsible for the signal level of FFOCT) and the resolution during wavefront optimization. Note that besides the gain in SNR and in spatial frequency energy,  fringes change their aspect becoming broaden and centered. 

%The adaptive-glasses approach can be easily implemented in other imaging modalities, as it does not require strict conjugation and MAL calibration (see \textcolor{blue}{Supplement 1} for results in wide-field microscopy).

%To demonstrate that our approach could also be used in other imaging modalities, we blocked the reference arm of the FFOCT, effectively converting it into a conventional  wide-field microscope. We then acquired images before and after aberration correction and computed their PSD as compared to the diffraction-limited PSD (Fig.~\ref{fig:UsafTarget}(c,d)). Aberrations induce loss of high spatial frequencies, blurring the image. The adaptive-glasses approach enables recovery of an almost diffraction-limited image of the USAF target after optimization.  

% old caption: 
%    \caption{Validation of the adaptive glasses approach on the USAF target. \textbf{a,c} USAF target image before and after aberration correction using FFOCT and wide-field microscopy respectively. Although the SD-OCT brightness was used as a merit function for both cases, the \textit{en-face} images could also be used, though requiring longer processing time. \textbf{b,d} PSD of USAF target image before and after aberration correction compared to the expected diffraction limit case using respectively the FFOCT and wide field microscope.}

\begin{figure}[H]
    \centering
    \includegraphics[width= \linewidth]{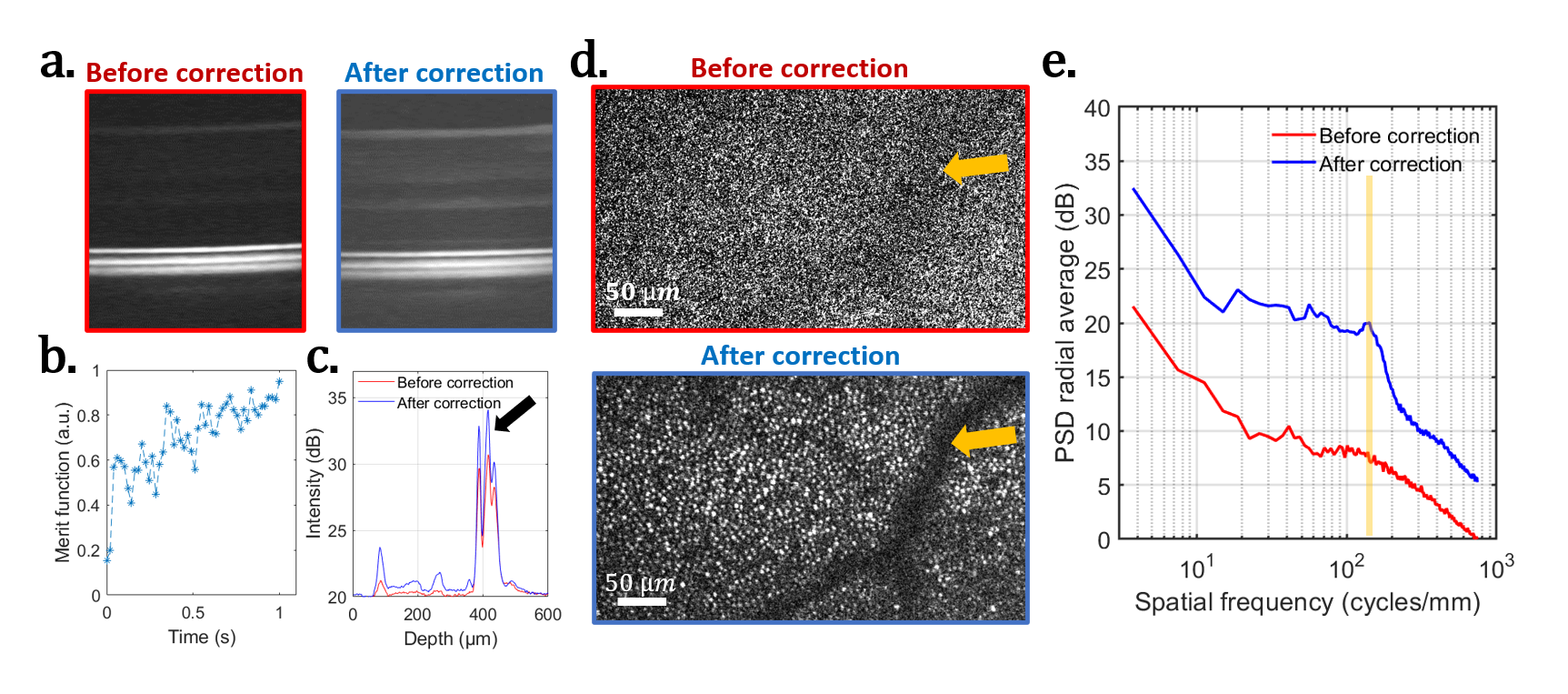}
    \caption{Validation of the adaptive glasses approach for \textit{in-vivo} retinal imaging. \textbf{a.} SD-OCT retinal cross-section before and after correcting for ocular aberrations. \textbf{b.} Values of the merit function during wavefront optimization using the DONE algorithm. \textbf{c.} Lateral average of SD-OCT B-scans highlighting the increased signal after aberration correction. Black arrow points to the retinal layer used for wavefront optimization. \textbf{d,e} Single non-averaged FFOCT frames before and after aberration correction, and their respective PSDs, outlining the gain in terms of SNR. Yellow arrow indicates the shadow of a vessel positioned in a retinal layer above the photoreceptor layer.}
    \label{fig:SignalGain}
\end{figure}

Figure~\ref{fig:SignalGain} shows the capacity of the adaptive-glasses approach to correct ocular aberrations and increase the SNR at the same time for the SD-OCT (Figs.~\ref{fig:SignalGain}(a,c)) and FFOCT (Figs.~\ref{fig:SignalGain}(d,e)) for \textit{in-vivo} retinal imaging. The wavefront optimization (Fig.~\ref{fig:SignalGain}(b)) was realized using the signal of the cone outer segment tips (COST - black arrow in Fig.~\ref{fig:SignalGain}(c)).  Figure~\ref{fig:SignalGain}(d) presents the same retinal zone (yellow arrow points to the shadow of a vessel) before and after correcting ocular aberrations with the adaptive glasses approach. A considerable increase in SNR is observed, thus making it possible to resolve the photoreceptor mosaic with a 7.5mm pupil diameter in a single non-averaged frame. PSDs of these images, highlighting the gain in terms of SNR, are given in Fig.~\ref{fig:SignalGain}(e), where the pale yellow column indicates the spatial frequency of the photoreceptor mosaic . 
%We computed a cone density of $17~463~cones/mm^2$ which is consistent with the literature \cite{curcio1990human}.

\begin{figure}[H]
    \centering
    \includegraphics[width= 1\linewidth]{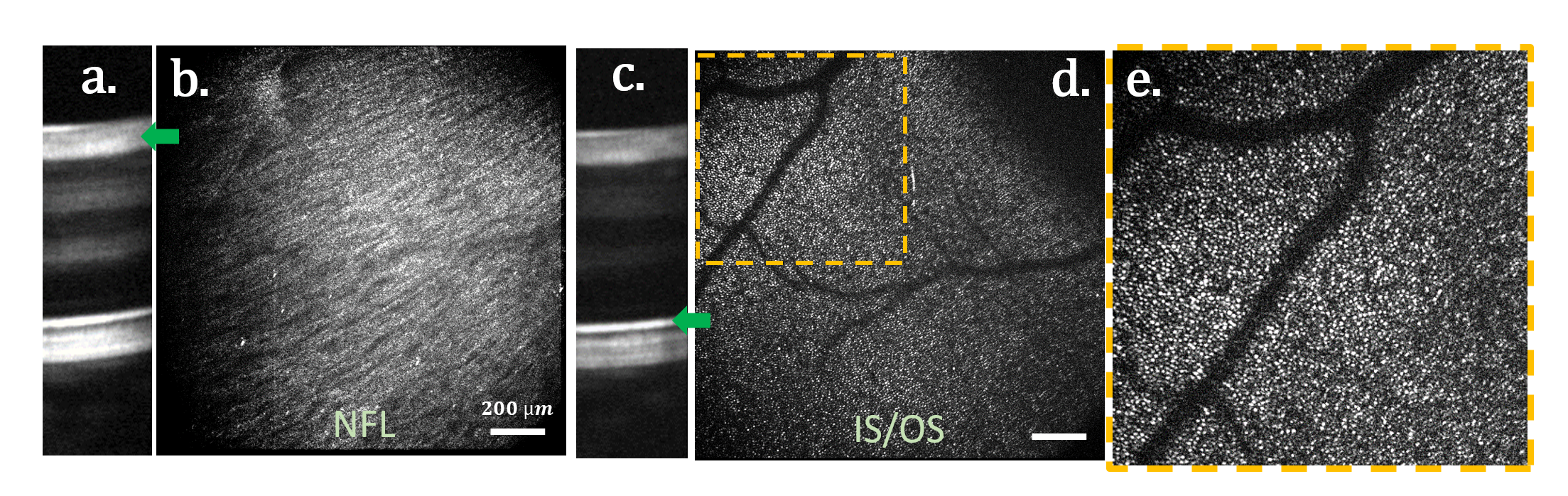}
    \caption{Wavefront optimization for a given retinal depth. \textbf{a,b} SD-OCT retinal cross-section and FF-OCT \textit{en-face} image acquired when optimizing for the retinal NFL (green arrow). \textbf{c,d} At the same region as \textbf{a,b} but at a different retinal depth after applying the wavefront optimization at the IS/OS junction. FFOCT images were acquired at $8\degree$ nasal with a $5\degree \times 5\degree$ FOV. \textbf{e.} Zoomed area of the photoreceptor mosaic from (\textbf{d}).}
    \label{fig:Depth}
\end{figure}

Using the SD-OCT and the axial motion stabilization, we can precisely position the FFOCT coherence gate at the retinal layer of interest. However, a mismatch of the coherence gate and focal plane positions produces low SNR FFOCT images. At the full-aperture, \textit{e.g.} for a 7-mm pupil diameter, the depth of focus is approximately ten times thinner than the retina \cite{Mece:20OIT}, making focus position an essential step. If the merit function did not take into account the retinal layer of interest, the wavefront optimization would be biased to the photoreceptor layer \cite{camino2020depth}, preventing high SNR FFOCT images of the inner retina. During each imaging session, SD-OCT B-scans are displayed in real-time, allowing the user to select the retinal layer of interest, where the coherence gate is then automatically positioned. In addition, when activating the wavefront optimization, only the brightness of the selected layer is taken into account, optimizing the focal plane position to the coherence gate position. Owing to this procedure, we were able to image nerve fiber layer (NFL) and photoreceptor inner/outer segment junction (IS/OS) at the same retinal region with ease (Fig.~\ref{fig:Depth}). Green arrows indicate the retinal layer selected for the merit function in the SD-OCT B-scan. \textcolor{blue}{Visualization 2} presents FFOCT images acquired at different depths in the NFL at $8\degree$ nasal, highlighting the high axial resolution afforded by FFOCT technique (\textit{i.e.} $8\mu m$).

Another benefit of correcting ocular aberrations for FFOCT is increased robustness. Indeed, FFOCT images are generated after a 2-phase demodulation step, meaning that they carry an amplitude signal modulated by the phase difference of two consecutive images, \textit{i.e.} $A\times cos(\Delta\phi)$ (where A is the amplitude signal, and $\Delta\phi$ the phase difference of two consecutive images) \cite{Mece_axial_stab_20}. Since phase modulation is performed almost randomly by the residual eye motion after correction \cite{Mece_axial_stab_20}, and ocular aberrations dampen the measured amplitude, the majority of acquired images are dominated by noise. Aberration correction restores the amplitude signal, thus increasing the number of high SNR images. \textcolor{blue}{Visualizations 3,4} presents an FFOCT image sequence after correcting ocular aberrations with the adaptive-glasses approach, where the photoreceptor mosaic and NFL respectively can be visualized in single frames and monitored over time with a 6ms resolution. 

\begin{figure}[H]
    \centering
    \includegraphics[width= \linewidth]{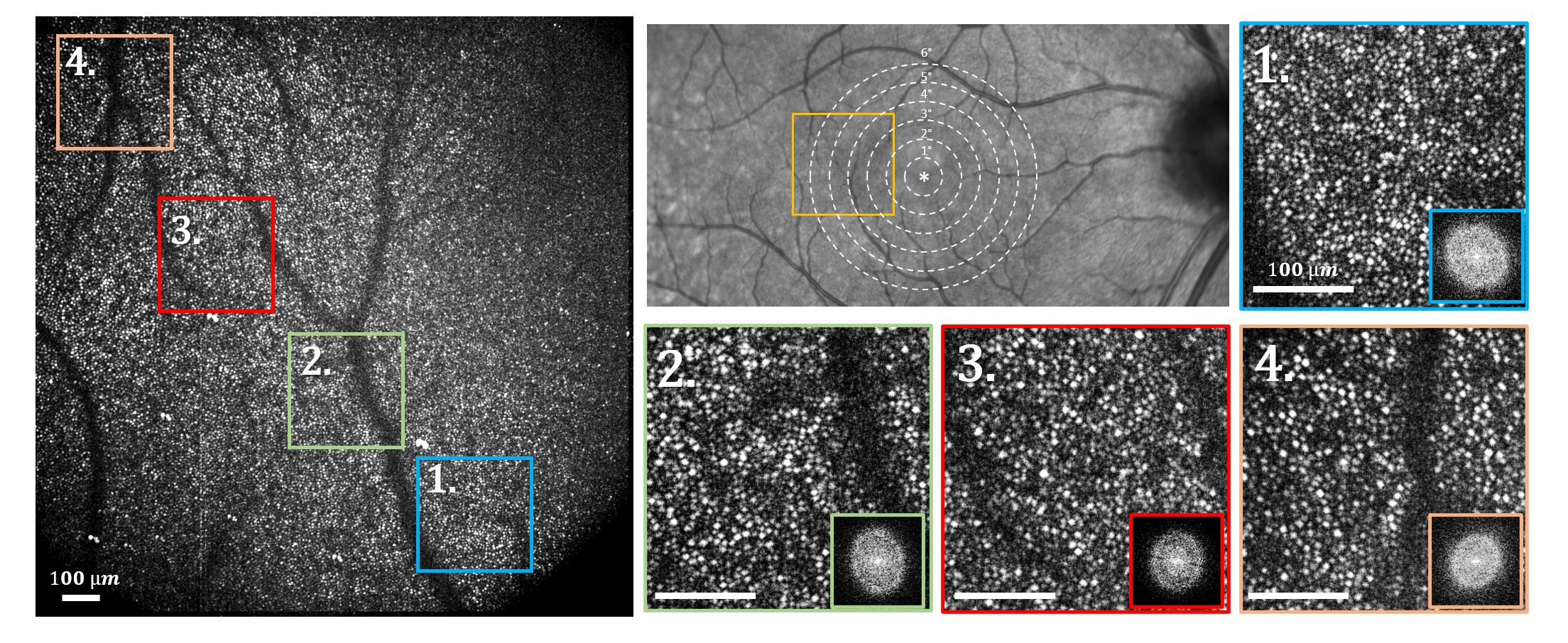}
    \caption{$5\degree \times 5\degree$ FOV retinal image as close as $2\degree$ from the foveal center, where photoreceptor can be resolved without any apparent isoplanatic patch limitation. Zoomed areas and their respective Fourier transforms are also presented.}
    \label{fig:WideFOV}
\end{figure}

 One important hurdle of AO-OCT for clinical translation is the challenge of allying high-spatial resolution with a wide FOV, which is beneficial for clinical applications. Indeed, the dense sampling of the scan required comes with a cost of low acquisition rate and image distortion due to eye motion. Moreover, aberration correction using AO is limited by the eye's isoplanatic patch to a small $2\degree \times 2\degree$ FOV. The combination of FFOCT and the adaptive-glasses approach opens a new avenue to wide FOV high-resolution retinal imaging in a compact imaging system (system footprint: $50 cm \times 30 cm$). %Additional advantages of FFOCT are that its lateral resolution is only weakly sensitive to symmetric aberrations, and it has a high-frame rate. The adaptive-glasses approach increases the FFOCT SNR and favors an anisoplanatic correction.
 Figure~\ref{fig:WideFOV} presents a $5\degree \times 5\degree$ FOV image obtained in a single shot (0.5s acquisition duration), as close as $2\degree$ from the foveal center, where photoreceptors can be resolved over almost the entire FOV (limited only  by the retinal curvature). Zoomed areas highlight that no apparent anisoplanatic effect is observed.

\begin{figure}[H]
    \centering
    \includegraphics[width= \linewidth]{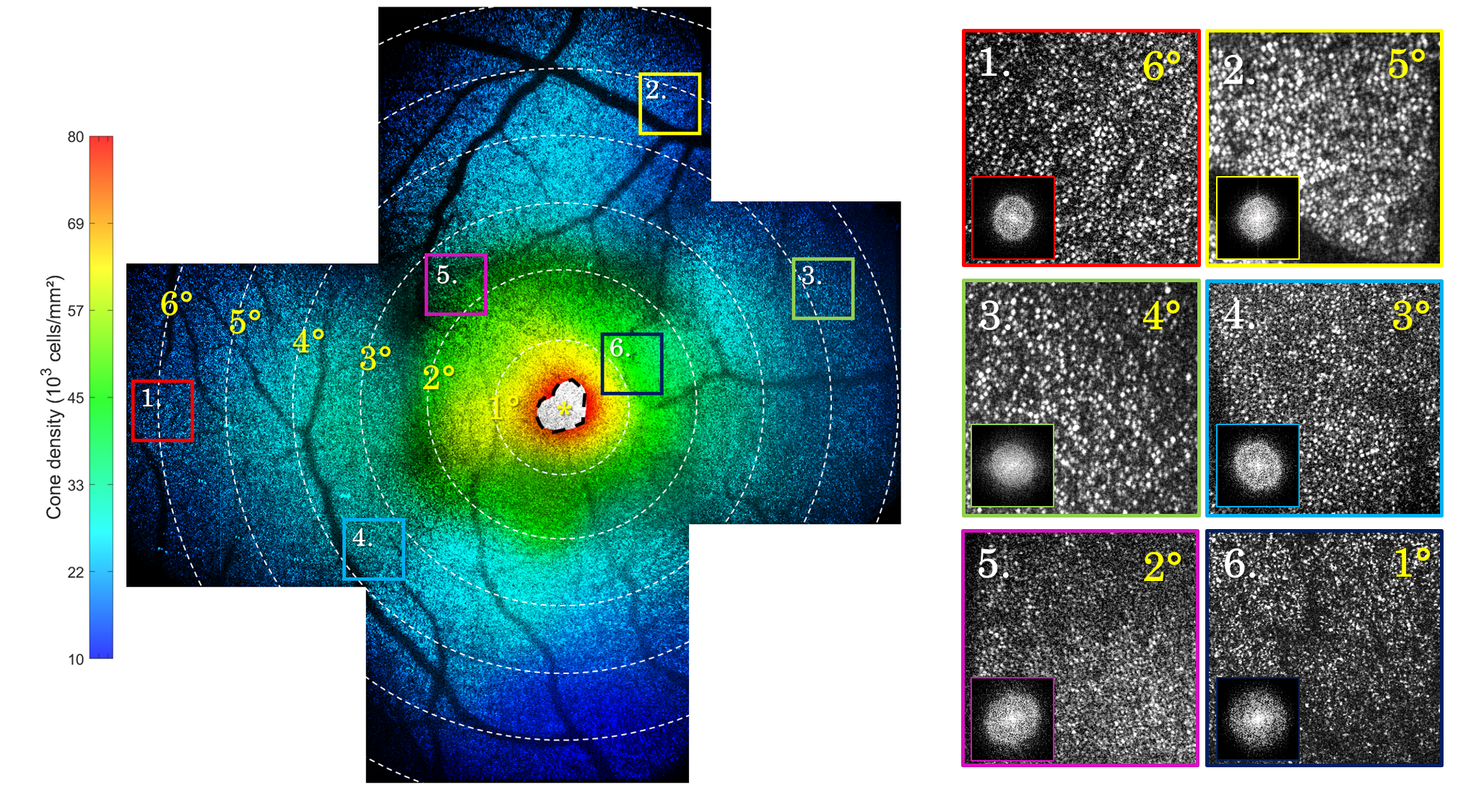}
    \caption{FFOCT retinal image over a $12\degree \times 12\degree$ FOV after stitching together five FFOCT images of $5\degree \times 5\degree$. The color map represents the computed photoreceptor density. Zoomed areas of $1\degree \times 1\degree$ FOV, chosen to be representative of different eccentricities, and their respective Fourier transforms. Black dashed area indicates where cone density was unreliably measured, but automatically detected and discarded.}
    \label{fig:Density}
\end{figure}

The wide FOV obtained in a single shot facilitates important tasks in the clinical environment to diagnose retinal disorders at early stages, such as image montaging and the computation of photoreceptor-based biomarkers  \cite{litts2017photoreceptor}. Figure \ref{fig:Density} shows a $12\degree \times 12\degree$ FOV image at the fovea composed of 5 images acquired at different retinal locations. Photoreceptor density is color coded and was consistent with the literature except within $0.5\degree$ of the fovea (black dashed area) where photoreceptors were not resolved \cite{curcio1990human} but were nevertheless automatically detected and discarded using the method proposed by \cite{cooper2019fully}. The total time necessary to obtain such an image (including subject alignment, image acquisition and processing) is about 15min. For comparison, an instrument with a $2\degree \times 2\degree$ FOV (\textit{i.e.}the typical size of the eye's isoplanatic patch) would need to stitch around 60 images to obtain the same image area with an image processing time multiplied by at least a factor of 10 \cite{Laslandes_multi_conjugate_AO_17,chui2008adaptive}.

\section{Conclusion}
We proposed the adaptive glasses approach as a wavefront sensorless AO method favoring small footprint, low optical complexity, and anisoplanatic correction, and extending the useful FOV over which high-resolution can be effectively achieved. We successfully applied this approach to \textit{in-vivo} retinal imaging, achieving 3D high-resolution images with $5\degree \times 5\degree$ FOV at 300Hz in a single shot. To the best of our knowledge, this is the first demonstration of AO successfully coupled to FFOCT for retinal imaging. Although we mainly illustrated the proposed approach for dual-channel SD-OCT and FFOCT retinal imaging, it can be adapted to other imaging modalities and samples. Finally, the combination of the adaptive glasses approach with FFOCT tackles those challenges that have so far prevented transfer of AO-OCT technology from bench to clinics. 
 
\section*{Funding}
HELMHOLTZ grant, European Research Council (ERC) (610110), IHU FOReSIGHT [ANR-18-IAHU-0001], Region Ile-De-France fund SESAME 4D-EYE [EX047007], French state fund CARNOT VOIR ET ENTENDRE [x16-CARN 0029-01], Fondation Visio grant.

\section*{Acknowledgments}
Michel Paques and José Sahel for their clinical expertise and support. Mathias Fink for technical support.

\section*{Disclosures}
The authors declare no conflicts of interest.

\bibliographystyle{ieeetr}
\bibliography{sample}

\end{document}